\documentclass[a4paper,fleqn]{cas-dc}

\usepackage[numbers]{natbib}
\usepackage{caption}

\def\tsc#1{\csdef{#1}{\textsc{\lowercase{#1}}\xspace}}
\tsc{WGM}
\tsc{QE}
\tsc{EP}
\tsc{PMS}
\tsc{BEC}
\tsc{DE}

\begin{document}
\let\WriteBookmarks\relax
\def\floatpagepagefraction{1}
\def\textpagefraction{.001}
\shorttitle{A Benchmarking Framework for Open-Source Ecosystem}
\shortauthors{Fenglin Bi et~al.}

\title [mode = title]{OpenPerf: A Benchmarking Framework for the Sustainable Development of the Open-Source Ecosystem}                      



\author[1]{Fenglin Bi}[style=chinese]

\credit{Conceptualization of this study, Methodology, Software}

\address[1]{School of Data Science and Engineering, East China Normal University, Shanghai, 200062}

\author[1]{Fanyu Han}[style=chinese]

\credit{Data curation, Writing - Original draft preparation}

\address[2]{School of Electronic and Information Engineering, Tongji University, Shanghai, 200070}

\author[2]{Shengyu Zhao}[style=chinese]

\author[1]{Jinlu Li}[style=chinese]

\author[1]{Yanbin Zhang}[style=chinese]

\author%
[1]
{Wei Wang}
\cormark[1]
\ead{wwang@dase.ecnu.edu.cn}
\cortext[cor1]{Corresponding author}


\begin{abstract}
Benchmarking involves designing scientific test methods, tools, and frameworks to quantitatively and comparably assess specific performance indicators of certain test subjects. With the development of artificial intelligence, AI benchmarking datasets such as ImageNet and DataPerf have gradually become consensus standards in both academic and industrial fields. However, constructing a benchmarking framework remains a significant challenge in the open-source domain due to the diverse range of data types, the wide array of research issues, and the intricate nature of collaboration networks. This paper introduces OpenPerf, a benchmarking framework designed for the sustainable development of the open-source ecosystem. This framework defines 9 task benchmarking tasks in the open-source research, encompassing 3 data types: time series, text, and graphics, and addresses 6 research problems including regression, classification, recommendation, ranking, network building, and anomaly detection. Based on the above tasks, we implemented 3 data science task benchmarks, 2 index-based benchmarks, and 1 standard benchmark. Notably, the index-based benchmarks have been adopted by the China Electronics Standardization Institute as evaluation criteria for open-source community governance. Additionally, we have developed a comprehensive toolkit for OpenPerf, which not only offers robust data management, tool integration, and user interface capabilities but also adopts a Benchmarking-as-a-Service (BaaS) model to serve academic institutions, industries, and foundations. Through its application in renowned companies and institutions such as Alibaba, Ant Group, and East China Normal University, we have validated OpenPerf's pivotal role in the healthy evolution of the open-source ecosystem.


\end{abstract}

\begin{graphicalabstract}
\includegraphics{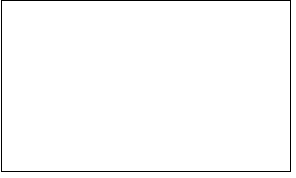}
\end{graphicalabstract}

\begin{highlights}
\item Introduced a novel benchmarking framework, tailored for the sustainable development of the open-source ecosystem, integrating data science task benchmarks, index benchmarks, and standard benchmarks.
\item Delineated nine data science benchmarking tasks spanning time series, text, and graph data, addressing six pivotal research domains.
\item Implemented three quintessential data science benchmarks and two index benchmarks, with the latter being recognized by the China Electronics Standardization Institute for open-source community governance evaluation.
\item Crafted a comprehensive open-source benchmarking suite, offering researchers and developers a modular and adaptive framework for diverse testing scenarios.
\item Transparently disseminated OpenPerf's benchmarking framework specifications as an open-source project and demonstrated its instrumental role in the sustainable growth of the open-source ecosystem through real-world case studies.
\end{highlights}


\maketitle

\section{Introduction}
Benchmarking\cite{zhan2022benchcouncil} is integral to computing, evident from leading supercomputers to current AI frameworks. As AI evolved, so did benchmarking standards, with datasets like ImageNet\cite{deng2009imagenet} and DataPerf\cite{mazumder2022dataperf} becoming pivotal. Accepted across academia and industry, these benchmarks offer objective evaluations for a range of AI innovations, from frameworks to tasks. Their influence has been profound, propelling both computer science and AI advancements\cite{ma2021world}.

In recent years, open-source software's surge has drawn significant global interest\cite{li2022saibench}. Globally, its pivotal role in digital innovation and transformation across varied organizations is acknowledged. Both academic and industrial researchers have delved into the open-source ecosystem using data-centric approaches. Their studies encompass corporate collaborations in open-source\cite{zhang2020companies}, the GitHub ecosystem's evolution\cite{constantinou2017socio, liao2019healthy}, developer contributions\cite{tsay2014let, rastogi2018relationship}, and textual analyses of developer comments\cite{venigalla2021understanding, huq2020developer, pletea2014security, sinha2016analyzing}. Open datasets in this realm have catalyzed such research, offering immense convenience and novel opportunities.

Despite numerous data tools and ongoing research yielding positive insights into the open-source software ecosystem, there remains a conspicuous absence of standardized benchmarks, annotations, and evaluation protocols\cite{kim2006study}. This gap has created a paradoxical landscape that is "data-rich, benchmark-poor." For precise insights into the evolution trends of open-source projects, community maturity, enterprise roles within the ecosystem, developer contributions, and project impact, a streamlined evaluation framework is essential\cite{ciolkowski2008towards, liao2018empirical}. Such a framework is not only vital for data consumers in the open-source realm but is also foundational for fostering a robust open-source software ecosystem. However, benchmarking within this domain is still in its embryonic stages. Given the diverse data sets, myriad research queries, and intricate nature of open-source collaborations, architecting an optimal benchmarking framework remains a formidable endeavor.

To address this gap in the open-source sector, this paper introduces OpenPerf—an innovative benchmarking framework tailored for the sustainable growth of the open-source ecosystem. Concurrently, we have crafted a flexible, extensible open-source benchmarking suite, offering a standardized framework for benchmarking within the open-source realm.

Employing this framework, researchers and developers can not only effortlessly assess the performance of open-source projects but also gauge key metrics like community activity and health from multiple dimensions. Our research provides fresh perspectives and tools for evaluating and nurturing the open-source sector.

The main contributions of this paper are:

\begin{enumerate}
\itemsep=0pt

\item Based on existing benchmarking frameworks\cite{zhan2021call}, we propose a benchmarking system designed for the sustainable evolution of the open-source ecosystem. This framework incorporates data science task benchmarks, index benchmarks, and standard benchmarks, aiming to provide benchmark references for various research directions within the open-source ecosystem.

\item For the designed framework architecture, we've defined 9 data science benchmarking tasks and implemented 3 representative data science benchmarks (open-source behavior data completion and prediction, OSS bot identification and classification, and open-source project recommendation based on link prediction), two index benchmarks (influence and activity), and one standard benchmark. Notably, the China Electronics Standardization Institute has adopted the activity and influence indices as evaluation criteria for open-source community governance\footnote{\url{https://www.cesa.cn/opinionDetail?id=310}}.

\item We've designed and implemented an open-source benchmarking suite which abstracts various components and stages of benchmarking, providing developers and researchers with a flexible and easily expandable framework. This ensures diverse testing scenarios and requirements can be swiftly accommodated.

\item OpenPerf's benchmarking framework specifications have been publicly released in the form of an open-source project. Additionally, we offer Benchmarking as a Service (BaaS) to various organizations, including the academic sector, industry, and foundations.
\end{enumerate}

\section{Related Work}

Benchmarking efforts have been initiated across numerous research areas, encompassing AI benchmarking datasets, test tasks, performance leaderboards, and more. Zhan et al.\cite{zhan2021call} elevated benchmarking to the level of "Benchmarking Science and Engineering," attempting to outline a systematic research framework and methodology. Deng et al.\cite{deng2009imagenet}, leveraging WordNet\cite{fellbaum1998wordnet}, introduced the ImageNet benchmark dataset, which offers certain advantages in terms of scale and diversity compared to other image datasets. Mazumderd et al.\cite{mazumder2022dataperf} proposed a machine learning dataset and a data-centric algorithm benchmarking suite (DataPerf). DataPerf can evaluate the quality of training and test data, covering a range of common machine learning tasks from vision, speech, acquisition, debugging, to dissemination. Hu et al.\cite{hu2020open} introduced the Open Graph Benchmark (OGB). They defined a unified evaluation protocol via data partitioning and evaluation metrics specific to applications, supporting various graph machine learning tasks from social information networks to biological networks, molecular graphs, and knowledge graphs. Zhou et al.\cite{zhou2022telegraph} presented a new benchmark dataset, TeleGraph, for graph link prediction. This dataset represents a highly sparse and layered information network with abundant node attributes, suitable for evaluating and advancing link prediction technologies. The aforementioned methods primarily centered around graph computation for their benchmark dataset research. Moreover, numerous researchers have proposed standard datasets for natural language processing. For instance, in sentiment analysis, Maria Pontiki et al.\cite{pontiki2016semeval} released datasets for the restaurant and e-commerce domains, aiming to discern users' sentiment polarities towards different aspects of entities. Based on this, they later added aspect term categorization and included a hotel domain dataset\cite{pontiki2015semeval}, providing more sub-tasks for sentiment analysis. A year later, they furnished datasets encompassing eight languages and seven distinct domains\cite{pontiki2016semeval}, allowing researchers to tackle diverse tasks due to the dataset's multifaceted nature. However, none of the existing efforts has culminated in a benchmarking dataset tailored for the sustainable development of the open-source ecosystem.

In recent years, with the adoption of open-source as a global digital development strategy, research in the open-source domain has garnered considerable attention from scholars both domestically and internationally. Numerous researchers have delved into the evolutionary processes of software ecosystems on GitHub. Hewapathirana et al.\cite{hewapathirana2017open} constructed a software ecosystem centered on open-source health information. Constantinou et al.\cite{constantinou2017socio} discerned that the evolution of open-source software ecosystems is heavily influenced by the vibrancy of its developer community, positing that both technological and social permanent shifts have profound impacts on the ecosystem. This observation underscores the continuous evolution and competitive nature of open-source communities. Research by Mockus et al.\cite{mockus2000case} suggests that core developers, serving as the linchpin and leaders of projects, generally contribute a significantly greater volume of code compared to peripheral developers. Additionally, Rastogi et al.\cite{rastogi2018relationship} asserted that while diversity and inclusivity are paramount to the sustained evolution of open-source software ecosystems, biases stemming from geographical locations and other demographic factors can potentially skew contribution evaluations, jeopardizing the health and collaborative spirit of the community.

In summary, most current benchmarking efforts revolve around a specific academic research direction, with a noticeable gap in benchmarking studies specifically aimed at the open-source ecosystem. Moreover, while a lot of research in the open-source domain delves into specific study points, most of them are predominantly data-analytic and lack concrete benchmarking standards. Given this backdrop, our paper integrates knowledge from the data science domain, existing datasets, and open-source application scenarios to propose a benchmarking framework for the open-source ecosystem. Built upon domain expertise, with benchmarking spaces as tasks and application scenarios as the orientation, we provide a comprehensive benchmarking architecture to facilitate the sustainable development of the open-source ecosystem.

\section{OpenPerf Architecture}

\subsection{Benchmarking framework for the Open-Source Domain}
One of the primary objectives of Benchmarking Science and Engineering\cite{zhan2021call} is to establish a standardized benchmarking hierarchy that spans multiple disciplines. This is achieved by introducing multidisciplinary benchmarks, standards, and evaluation metrics to communicate the latest techniques and practices in benchmarking science and engineering.

A pivotal aspect of benchmarking is maintaining consistency in benchmarks, which can generally be achieved by: (1) Uniformly defining measurement standards and units of measurement; (2) Implementing measurement standards and units with varying precision levels; and (3) Traceability and calibration of the benchmarking framework. Traceability is a characteristic of the measurement result, linking it to a reference through a complete, recordable calibration chain, with each link in the chain potentially influencing the measurement result. Calibration is the process of comparing the results of an entity, whose measurements are unknown or unchecked, with those of an entity that has known characteristics.

\begin{figure}
	\centering
		\includegraphics[scale=.15]{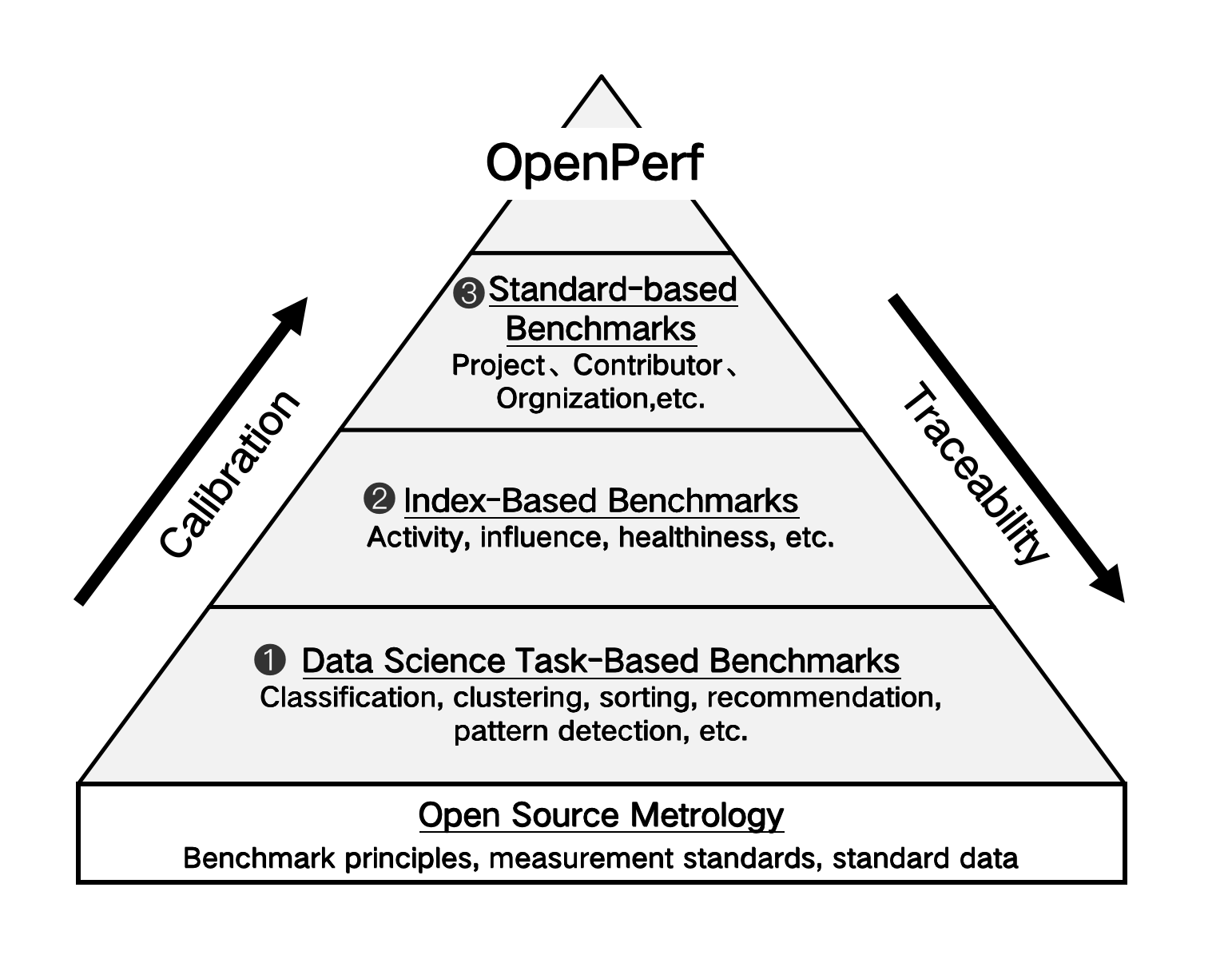}
	\caption{Benchmarking for the open-source}
	\label{FIG:1}
\end{figure}

Benchmark tests can be broadly categorized into the following types\cite{zhan2021call}: computing measurement standards (e.g., LINPAC, TPC), representative workloads (e.g., MLPerf), data science standard tasks and standard datasets (e.g., ImageNet, OGB), representative datasets (like various indices in the financial sector), and best practice benchmarks (e.g., best practices in various industries/business domains).

Based on the aforementioned benchmark categories and the unique application scenarios in the open-source domain, this paper proposes a hierarchical benchmarking framework tailored for open-source software development and ecosystem evolution, as illustrated in Figure~\ref{FIG:1}. In this context, we collectively refer to benchmarking principles, measurement standards, and standard data in the open-source domain as Open-source Metrology. The introduction of benchmarks requires the support of standard data and the determination of measurement standards. Standard data refers to datasets that have clearly defined benchmarks based on specific research directions. Such data can include predefined labels; for instance, ImageNet represents a standard dataset. Measurement standards pertain to the measurable properties of the subject being measured, like length or height. In machine learning, evaluation metrics are analogous to measurement standards.

\subsection{Data Science Task-Based Benchmarks}
In complex computational domains, benchmarking is a critical assessment of performance. The first category of the OpenPerf framework—data science task-based benchmarks—focuses on standard tasks, such as classification, clustering, and ranking. The core advantage of such benchmarks is that they provide academic researchers with a framework specific to their field of study, allowing them to evaluate and compare on a unified platform. Figure 2 provides a detailed depiction of the construction process of such benchmarking tasks.

\begin{figure}
	\centering
		\includegraphics[scale=.15]{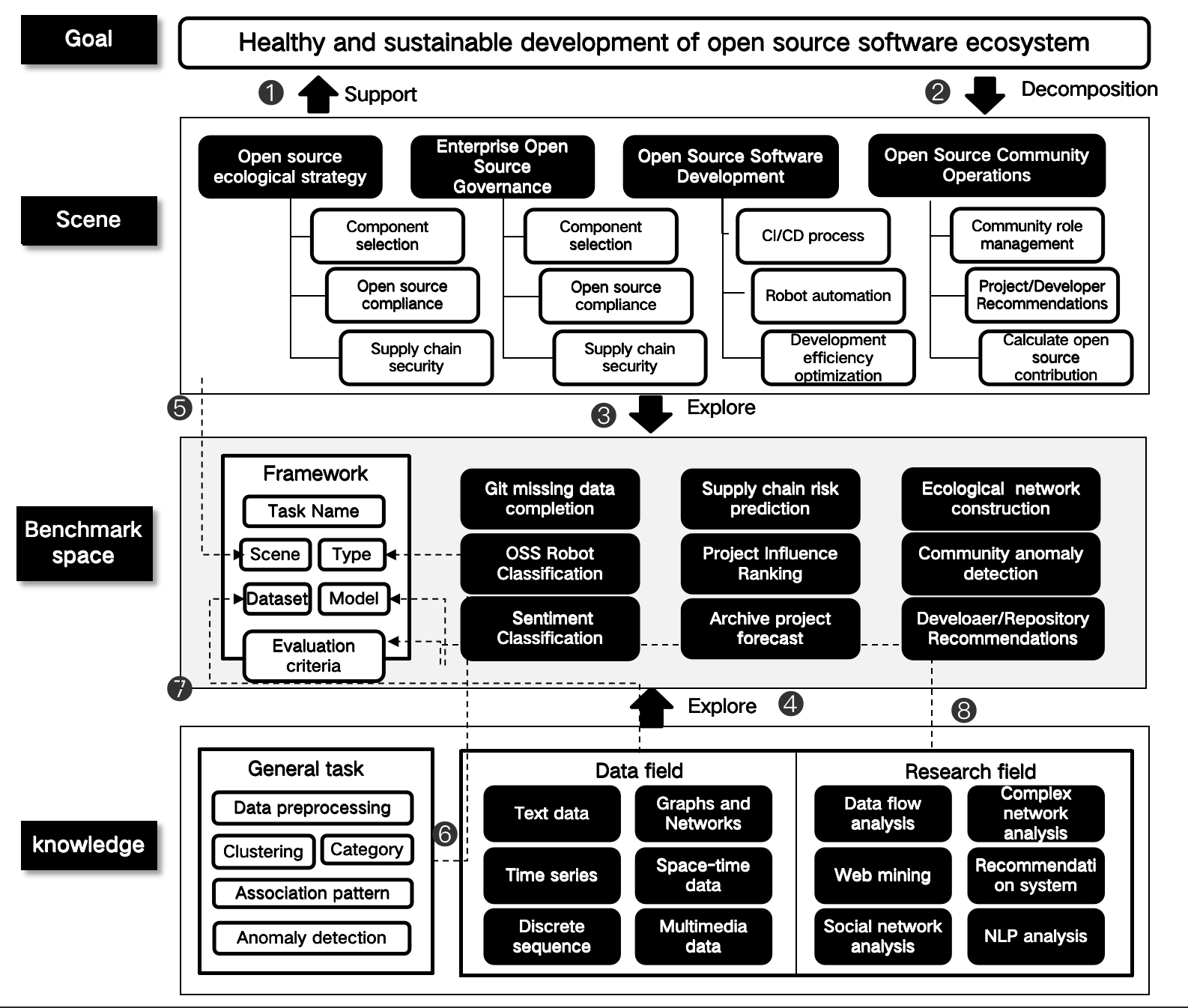}
	\caption{Process of Constructing Data Science Benchmarking Tasks}
	\label{FIG:2}
\end{figure}

In the vast realm of benchmarking, each specific benchmark is meticulously designed to capture a set of specific characteristics and performance metrics. Within OpenPerf's data science task-based benchmarks, each benchmark comprises the following six core elements:

\begin{itemize}
\item Task Name: Clearly describes the benchmark's primary purpose and focus, for instance, "Open-source Automation Robot Identification and Classification."
\item Open-source Scenario: Provides the practical application context for the benchmark, showcasing its real-world value and applicability, like Open-Source Community role management.
\item Task Type: The benchmark task can be either specific or abstract, ranging from classification to log data classification. 
\item Dataset: Refers to the benchmark dataset corresponding to the task.
\item Model Design: This not only provides predefined benchmarks but also offers researchers opportunities to compare with existing models.
\item Evaluation Metrics: To ensure consistency and accuracy in assessment, each benchmark clearly outlines a set of evaluation metrics, like Accuracy, Precision/Recall, F1-score, and AUC.
\end{itemize}

Furthermore, drawing from the knowledge framework of data analysis and mining\cite{azzalini2012data}, tasks in the open-source domain are categorized based on different perspectives:

\begin{itemize}
\itemsep=0pt
\item Task Perspective: Data preprocessing, clustering, classification, association patterns, anomaly analysis.  
\item Data Perspective: Text, time series, discrete sequence, graphs and networks, multimedia, spatio-temporal.
\item Research Perspective: Data stream analysis, complex network analysis, web mining, social network analysis, NLP analysis, recommendation systems.
\end{itemize}

\subsection{Index-Based Benchmarks}
With the growing demand for performance evaluation in the industrial sector, index-based benchmarks are emerging as principal evaluation tools aimed primarily at practical application domains. Compared to traditional data science task-based benchmarks, they offer more refined and specific avenues for evaluation.

The essence of index-based benchmarks lies in their role as "benchmarking units." The unitary nature of such designs enables deep, accurate measurements of specific attributes of a given subject. For example, when industrial enterprises aim to assess the community activity of their open-source projects, the market impact of their projects, or the overall health of their open-source initiatives, they can refer to the corresponding index-based benchmarks. These benchmarks provide enterprises with a quantified approach, allowing for deep insights and analysis of their operations.

The attributes, namely activity, influence, and health, serve as representative examples of what index-based benchmarks can encompass. As the dynamics of industrial requirements shift, the diversity and application domains of such benchmarks correspondingly expand, affording enterprises an increasingly nuanced and specific set of evaluative criteria.

Index-based benchmarks offer the industrial sector a practical, clearly directive evaluation tool, ensuring that enterprises make informed, substantiated decisions.

\subsection{Standard-Based Benchmarks}
Within the diverse domain of benchmark testing, standard-based benchmarks stand distinct, representing the industry's best practices and pinnacle achievements. They are not merely simple evaluation criteria but serve as an emblem of exceptional industry standards, providing other entities with a clear objective and reference to aspire towards.

A defining characteristic of these benchmarks is their measurability. They establish a clearly defined performance level, recognized and endorsed by the broader industry. This means that when an entity attains or surpasses this level, it is deemed a frontrunner in its field, setting a standard for others to learn from and emulate.

The core objective of this type of benchmark testing is to learn from the best and strive for excellence. This philosophy is equally applicable in the open-source digital ecosystem. We observe standard-based open-source projects that are widely acknowledged by the vast community due to their superior quality, extensive user base, and consistent contributions. Furthermore, standard-setting open-source developers, with their relentless dedication and ongoing contributions to open-source projects, emerge as community leaders. Likewise, leading open-source enterprises, through their proactive engagement and significant contributions to the open-source ecosystem, have garnered industry-wide respect and acknowledgment.

\subsection{Challenges of the OpenPerf Benchmarking Framework}
The OpenPerf benchmarking framework encompasses multiple facets and dimensions within the open-source domain. From the wide range of task types and research orientations in Data Science Task-Based Benchmarks, the unique measurement units in Index-Based Benchmarks, to the exceptional performance standards in Standard-Based Benchmarks, each benchmark category presents its distinct applications and challenges.

Data Science Task-Based Benchmarks:(1)Diversity of Research Orientations: Given the multitude of research pathways in data science, crafting a universal benchmark poses a challenge.(2)Standardization: Developing a standardized assessment tool that offers consistent, objective evaluations across varied research methodologies is a colossal undertaking\cite{grunzke2017challenges}.

Index-Based Benchmarks:(1)Metric Acceptance: The primary challenge here is introducing and garnering widespread acceptance for specific measurement units within the industry. These units must not only accurately capture attributes but also require extensive validation.(2)Dynamic Adaptability: As the nature of entities and their attributes evolve, benchmarks must be adjusted to maintain their relevance\cite{bouckaert1993measurement}. 

Standard-Based Benchmarks:(1)Dependency on Index-Based Benchmarks: These standards are fundamentally linked to the precise definition and measurement of Index-Based Benchmarks. Consequently, their efficacy is tightly coupled with the accuracy of the Index-Based Benchmarks.(2)Defining Excellence: Setting a benchmark for excellence necessitates a profound understanding of the domain and foresight into its future developments\cite{antonaras2009strategic}.

The grand vision of the OpenPerf benchmarking framework is to construct a comprehensive and systematic evaluation tool, aimed at encompassing various benchmarking tasks within the open-source ecosystem. While current accomplishments are promising, further in-depth research is required in the future to refine and detail the definitions and applications of each subcategory benchmark, catering to the continuous evolution and changes in the open-source community.

The pursuit of the OpenPerf benchmarking framework is to establish a comprehensive, in-depth, and continuously optimized assessment tool within the open-source domain, laying a solid foundation for future research and applications.

\subsection{OpenPerf Suite Tools}

\begin{figure}
	\centering
		\includegraphics[scale=.15]{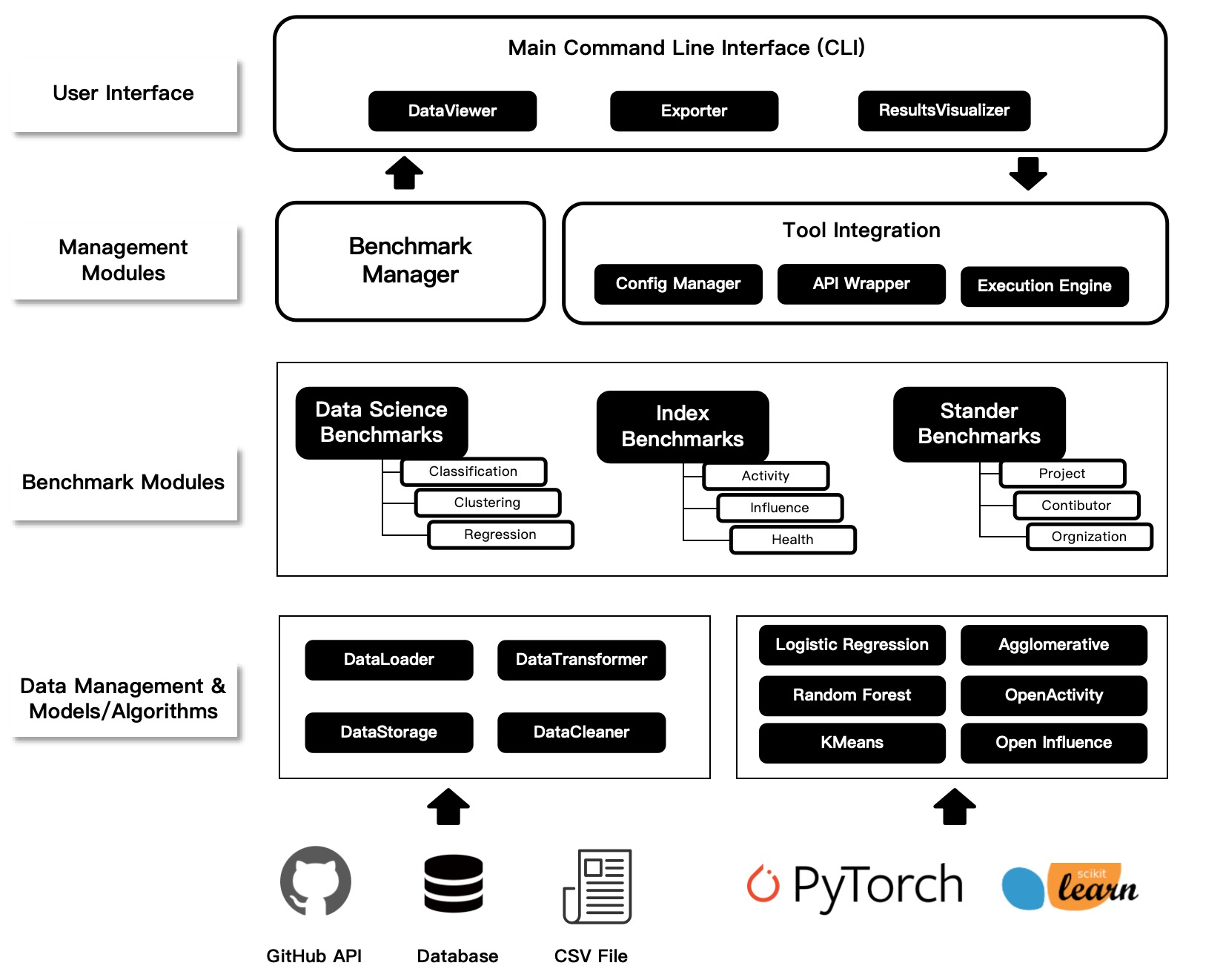}
	\caption{OpenPerf Suite Tool Architecture}
	\label{FIG:3}
\end{figure}

The OpenPerf Suite offers a cutting-edge architecture meticulously designed to furnish the open-source community with a comprehensive benchmarking toolset. As depicted in the architecture, the design is a result of rigorous research, ensuring that it caters to a myriad of benchmarking needs while seamlessly integrating with a variety of tools.

\begin{itemize}
\item User Interface: The top layer provides a user interface, serving as a direct point of interaction for users with the OpenPerf suite. Primarily driven by the main Command-Line Interface (CLI), this tier offers users access points to the myriad functionalities within OpenPerf. Components like DataViewer, Exporter, and ResultsVisualizer ensure an interactive experience, delivering seamless capabilities to view and export benchmarking results.

\item Management Module: The Management Module stands as the backbone of the suite, orchestrating and controlling various operations. The Benchmark Manager, pivotal in this layer, coordinates the entire benchmarking process. Simultaneously, under the tool integration domain, entities like Config Manager, API Wrapper, and Execution Engine guarantee smooth integration with numerous external tools, highlighting the suite's distinction in scalability and interoperability.

\item Benchmarking Module: Delving deeper, the benchmarking module encapsulates the essence of the benchmarking logic. Divided into three distinct categories—Data Science Benchmarks, Index Benchmarks, and Standard Benchmarks—each represents a specific domain within open-source projects. This design decision, rooted in extensive academic literature and industry needs, ensures the OpenPerf suite maintains both comprehensiveness and agility.

\item Data Management and Model/Algorithm Layer: At the data management and model/algorithm level, components like DataLoader, DataTransformer, DataStorage, and DataCleaner collaborate to handle and manage data efficiently. Concurrently, various algorithms, spanning from conventional machine learning models to domain-specific algorithms, manifest the suite's commitment to offering avant-garde solutions.

\item Data Sources and Libraries: At the foundation, diverse data sources like the GitHub API, databases, and CSV files exhibit the suite's adaptability in acquiring and processing varied data types. The organic integration with renowned libraries such as PyTorch and Scikit-learn further exemplifies OpenPerf's commitment to staying pertinent and harnessing best practices within the domain.
\end{itemize}

The architecture of OpenPerf embodies a harmonious blend of stringent academic research and real-world application needs. A prominent research challenge at its inception was striking a balance between generality (to cater to a wider audience) and specificity (to deliver value to specific domains). Given the dynamic nature of the open-source community, OpenPerf's modular design ensures adaptability. Future research could delve into more refined benchmarks.

\subsection{OpenPerf Benchmark Service Architecture}

\begin{figure}
	\centering
		\includegraphics[scale=.15]{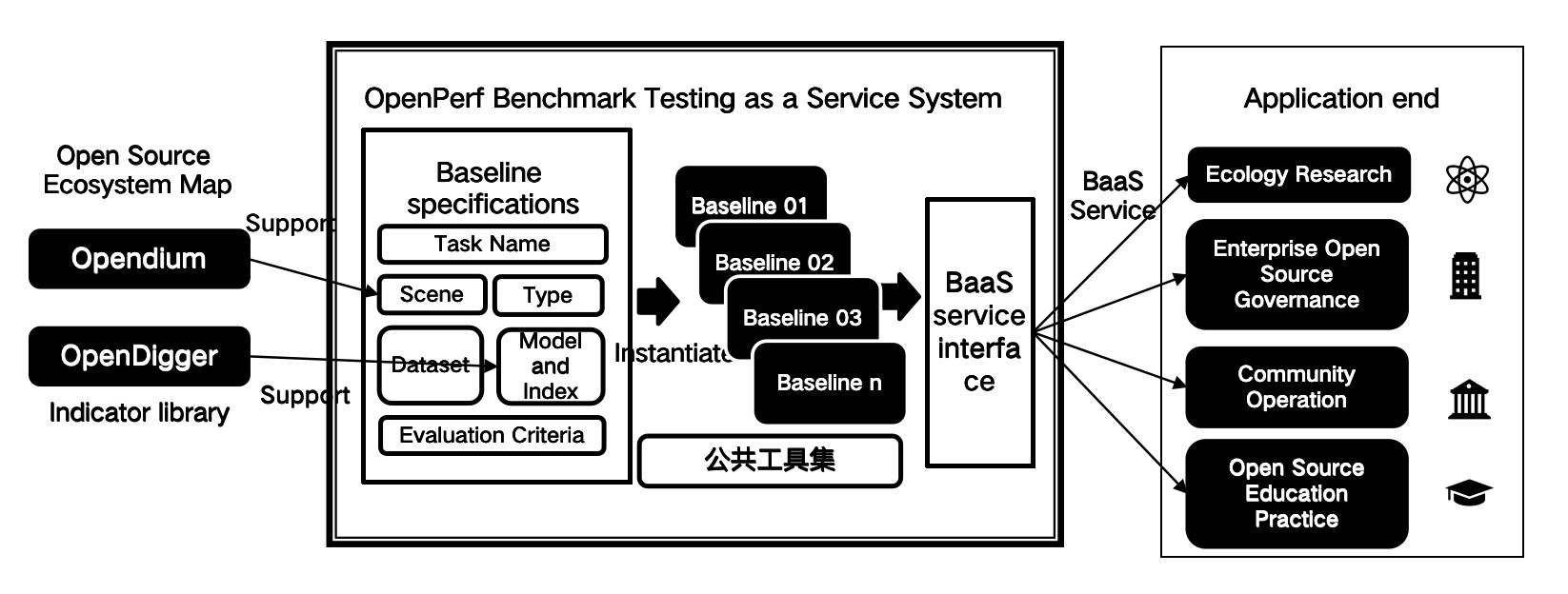}
	\caption{OpenPerf Benchmark-as-a-Service Architecture}
	\label{FIG:4}
\end{figure}

Building upon the aforementioned benchmarking framework, this article introduces a Benchmark-as-a-Service (BaaS) system tailored for the sustainable development of the open-source ecosystem. With the primary aim of fostering a healthy and sustainable growth of the open-source software ecosystem, this architecture comprises four pivotal components: benchmark specifications, benchmark instances, a public toolset, and BaaS service interfaces, designed to cater to diverse application scenarios for end-users.
\begin{itemize}
\item Benchmark Specifications: This provides a unified framework to define each benchmark instance within OpenPerf. A standardized benchmark instance encompasses six core components: task name, application scenario, task type, dataset, algorithm model, and evaluation standards.

\item Benchmark Instances: Represents a collection of specific benchmark instances, such as open-source behavioral data completion and prediction benchmarks, open-source automation robot identification and classification benchmarks, and link prediction-based open-source project recommendation benchmarks, among others.

\item Public Toolset: Furnishes a range of universal tools, including data loading tools, result evaluation tools, and leaderboards.

\item BaaS Service Interface: Offers final data services to various applications, spanning scenarios like open-source research in universities, corporate open-source governance, open-source community operations, and open-source educational practices.
\end{itemize}  

In conjunction, the architecture is supplemented by two external support projects, aiding researchers and designers in developing distinct benchmark instances:

\begin{itemize}
\item OpenDigger\footnote{\url{https://github.com/X-lab2017/open-digger}}: An open-source initiative focusing on data collection and analysis infrastructure for the open-source ecosystem, primed to provide various data sources for OpenPerf.

\item Opendium\footnote{\url{https://github.com/X-lab2017/open-research/tree/main/Opendium}}: An open-source compendium (dictionary) project charting the knowledge graph of the open-source ecosystem. It meticulously categorizes open-source knowledge essential for benchmark instance design, encompassing application scenario classification, task type classification, dataset classification, and data science domain classification.
\end{itemize}

The service architecture encompasses four principal application scenarios in the open-source domain:

\begin{itemize}
\item Open-source Ecosystem Research: Encompasses analyses and understandings of the overall state of the open-source ecosystem from economic, social, educational, academic, policy, legal, and cultural dimensions, such as demographic information, supply chain identification and situation, and technological ecosystem evolution.

\item Corporate Open-source Governance: Refers to various business scenarios encountered by the corporate Open-source Program Office (OSPO), like open-source component selection, open-source compliance, supply chain security, ranking, and incentives.
\end{itemize}

\section{Data Science Task}

\begin{table*}
\small
\centering
\captionsetup{justification=centering}
\caption{Benchmarking Tasks}\label{tbl1}
\begin{tabular}{ccccc}
\toprule
Benchmarking Task & Data Type & Problem Type & Scene & Research Field\\
\midrule
 \begin{tabular}[c]{@{}c@{}}Behavior Data Completion \\ and Prediction\cite{chen2023temporal}\end{tabular} & Time Series & Regression & Enterprise Governance & Data Flow \\
 \begin{tabular}[c]{@{}c@{}}OSS Bot Identification \\ and Classification\cite{bi2023bothawk}\end{tabular} & Time Series & Classification & Software Development & Data Flow \\
  \begin{tabular}[c]{@{}c@{}}Community Sentiment \\ Classification\cite{venigalla2021understanding, you2022ask} \end{tabular} & Text Data & Classification & Community Operations & NLP \\
 \begin{tabular}[c]{@{}c@{}}Software Supply Chain \\ Risk Prediction\cite{librantz2021risk}\end{tabular} & Time Series & Regression & Ecosystem Strategy & Complex Network \\
 Project Influence Ranking\cite{zhao2022} & Graph \& Network & Ranking & Community Operations & Complex Network \\
 Archived Project Prediction\cite{xia2023understanding} & Time Series & Regression & Enterprise Governance & Web Mining \\
 Network Metric Prediction\cite{xia2022exploring} & Graph \& Network & Regression & Enterprise Governance & Data Flow \\
 \begin{tabular}[c]{@{}c@{}}Community Anomalous \\ Detection\cite{chen2021celof}\end{tabular} & Time Series & Anomaly Detection & Enterprise Governance & Complex Network \\
 Project Recommendation\cite{wang2022} & Graph \& Network & Recommendation & Community Operations & Recommendation \\
\bottomrule
\end{tabular}
\end{table*}

In this chapter, drawing on the research results of our team, we describe nine of the most representative tasks (as shown in Table 1) and introduce the background, problem types involved, and their specific applications in the open-source domain for each of these tasks.

\subsection{Behavioral Data Completion and Prediction} 
In the domain of open-source software, high-quality data governance has become a pivotal factor propelling open-source development. Especially in this era of accelerated digital transformation, data has transformed into a vital resource. High-quality data aids in accurately capturing the current state of the entire project, while low-quality or even missing data might lead to errors in research conclusions, subsequently affecting decision-making for open-source projects. When collecting developer behavioral data from open-source platforms like GitHub, some open-source project behavioral activity data cannot be fully gathered due to reasons like the platform's internal mechanisms, API restrictions, collection technologies, and fluctuations in internal services. This results in missing behavioral data, posing significant impediments to subsequent granular research\cite{vasilescu2015data}.

The key focus of the open-source behavioral data completion and prediction task is to design a model that can both fill in the missing values of open-source behavioral data and predict future trends. The model needs to adeptly mine the periodicity and correlations of open-source behavioral data, retain timestamp information of the data, and not rely on the data's prior information and probability distribution. Moreover, the model should effectively handle situations of data omission, such as GitHub's collection limitations and fluctuations in network services.

\subsection{OSS Bot Identification and Classification}

Collaboration, as a societal phenomenon, is increasingly pivotal in the software development lifecycle. Popular social coding platforms, such as GitHub, Bitbucket, and GitLab, offer environments for developers to share workspaces. However, large-scale collaboration introduces significant workloads for repository maintainers. They are tasked with communicating with contributors, reviewing source code, addressing contributor license agreement issues, clarifying project guidelines, executing tests, building code, merging pull requests, and more. To alleviate the burden of these repetitive tasks, open-source software projects have recently begun to harness various software bots to streamline their operations. However, the use of bots has introduced a series of challenges, including impersonation, spam, bias, and security risks\cite{shvets2018automatic}.

The task of OSS Bot Identification and Classification necessitates the development of a model capable of identifying and classifying bot behaviors within open-source software projects. This model should efficiently categorize based on the bots' behavioral patterns and objectives. Beyond high predictive accuracy, the model must also exhibit robust explainability and sound rationale.
In the development of open-source Software, developers' behavioral data exhibits intricate correlational and cyclical characteristics. These traits hold significant value for understanding developers' behavioral patterns, project progression, and the implications for software quality. However, the complexity of these traits renders traditional statistical methods inadequate for accurately predicting and analyzing developers' behaviors. Open-source operators are particularly interested in leveraging statistical indices to enhance network metrics.

The prediction task for open-source network metrics predominantly employs statistical indices to fit network metrics. This approach necessitates adaptability to the unique characteristics of OSS data, encompassing the intricate interrelations among various behavioral data and the cyclical nature of developer behaviors.

\subsection{ Community Sentiment Classification}

Sentiment analysis tasks based on comments in the open-source domain are relatively limited compared to popular domains like restaurants and e-commerce. The sentiment of developers can significantly influence task quality, productivity, creativity, team harmony, and job satisfaction\cite{rishi2017affective, kaur2022analysis}. Analyzing sentiments from comments provides insights into developers' behaviors and opinions about specific aspects of a project. This is crucial for the healthy evolution of a project and can also contribute to improving developer work efficiency.

Most comments on GitHub are neutral. How can one extract comments that contain developers' perspectives from a vast amount of text? Different types of comment texts might convey varying opinions. Moreover, developers often express views about different facets of the open-source community. Extracting granular, specific sentiments from these comments remains a significant challenge for this task.

\subsection{Software Supply Chain Risk Prediction}
Open-source software, emblematic of the cornerstone of information technology innovation, has witnessed an unprecedented escalation in the intricacy and magnitude of its dependencies. This, in turn, has led to an exponential rise in associated security vulnerabilities. Such challenges have the potential to trigger cascading repercussions throughout the open-ource software supply chain, affecting every subsequent system built upon it\cite{ohm2020backstabber}. At present, the global appetite for open-source components is rapidly accelerating; however, the security posture of its supply chain is witnessing a concurrent decline. It's worth noting that, given the significance of the open-source software supply chain in geopolitical strategic dialogues and as a bedrock for enterprise operations, its integrity and security are of utmost importance.

Efforts towards predicting risks in the software supply chain mandate the comprehensive quantification of an array of risk determinants. These span intrinsic software characteristics such as lines of code, foundational architecture, and security loopholes, as well as extrinsic facets like the composition and dynamics of development teams, engagement metrics, and overall traction within the developer community. There is a pressing need to architect and enforce predictive models proficient in quantifying these risks. Such endeavors aim to empower organizations with the insights required for optimized supply chain governance.

\subsection{Project Influence Ranking}

Open-source projects not only offer mature and reliable code, saving developers substantial amounts of time, but also present vast platforms for innovation and collaboration. Whether in the technological sector or in areas such as education, government, and non-profit organizations, open-source projects play an indispensable role. This trend underscores the importance of assessing the influence of open-source projects\cite{singh2010small}. However, the influence of open-source projects is contingent upon multiple factors, including but not limited to project activity, community size, number of contributors, code quality, and project stability. Appropriately weighing and integrating these factors is crucial in evaluating the influence of open-source projects. Such an assessment can aid developers and organizations in identifying the most valuable projects deserving time and resources and can also offer insights for the optimization and improvement of these projects.

The task of ranking the influence of open-source projects should not only gauge the overall influence of a project but also precisely quantify each developer's contribution within that project. This would facilitate the provision of corresponding incentives, fostering their active participation in the open-source realm.

\subsection{Archived Project Prediction}

The sustainability of the open-source software ecosystem has emerged as a pivotal topic. Unlike traditional software development lifecycle models, which have clear delivery goals, dedicated teams, and milestones, open-source software development relies heavily on self-organized contributors and voluntary work in its early stages. This dynamic has led to the archiving of open-source software projects. For developers active on GitHub and others who monitor specific open-source projects, predicting whether a project might be archived is of utmost significance\cite{xia2023understanding}.

The task of predicting open-source project archiving necessitates the analysis of a vast amount of data. This encompasses, but is not limited to, the project's commit history, its activity levels, maintainer information, and project maturity, among other factors. Through this data, the task aims to identify key factors that might lead to a project's archiving.

\subsection{Network Metric Prediction}

In open-source software development, developers' behavioral data exhibits intricate correlational and cyclical characteristics. These traits hold significant value for understanding developers' behavioral patterns, project progression, and the implications for software quality. However, the complexity of these traits renders traditional statistical methods inadequate for accurately predicting and analyzing developers' behaviors. Open-source operators are particularly interested in leveraging statistical indices to enhance network metrics.

The prediction task for open-source network metrics predominantly employs statistical indices to fit network metrics, which necessitates adaptability to the characteristics of OSS data, encompassing the intricate interrelations among various behavioral data and the cyclical of developer behaviors.

\subsection{Community Anomalous Detection}

Behavioral data from developers in open-source projects is a valuable information asset. It allows for the monitoring of project dynamics, timely identification, and handling of anomalous behaviors, and facilitates the optimization of project management protocols\cite{al2022cyber}. As the number of project participants grows, manually monitoring all anomalous behaviors becomes impractical, necessitating a rapid and efficient automated monitoring solution.

The anomaly detection task in open-source community behavior aims to implement a real-time anomaly detection framework. This framework should continuously monitor the vast streams of developer behavior data for anomalies. The objectives are multifaceted: to provide early warnings for anomalies, similar to events like those concerning the NPM library, and timely feedback for community managers to adjust project management protocols; to reduce the computational scale and time taken by the algorithm, minimizing infrastructure demands; and to filter out anomalous behavior data, paving the way for finer-grained research on the data in subsequent stages.

\subsection{Project Recommendation Based on Link Prediction}

Open-source software development encourages developers worldwide to participate in development tasks, such as bug fixes, code testing, and documentation improvements\cite{xia2013accurate}. Each developer has their unique project needs and technical backgrounds. Failing to match them with suitable open-ource projects can hinder the smooth progress of development tasks, negatively impacting the growth of individual open-ource projects and even the broader open-ource ecosystem.

Recommendations based on link prediction primarily aim to suggest projects of interest to developers and community operators by mining relationships between projects. Recommending intriguing open-ource projects to developers can help them comprehensively understand their target technical fields, minimizing the wastage of developmental resources.

\section{Representative Task-Based Benchmark Results}

In this section, we select three representative task-based benchmark tests - Behavior Data Completion and Prediction, OSS Bot Identification and Classification, and Project Recommendation based on Link Prediction. We provide datasets, evaluation metrics, and evaluation results, offering benchmark reference samples for subsequent researchers.

\subsection{Behavior Data Completion and Prediction}

\begin{table*}
\centering
\captionsetup{justification=centering}
\caption{The prediction results of eight algorithms on five groups of OSS behavior datasets}\label{tbl2}
\begin{tabular}{@{} llllllllll@{} }
\toprule
Dataset & Metric & TAMF & TRMF & L-SVR & L-R & KNN & iForest & K-fold & R-chain \\
\midrule
Pytorch & NMSE & 0.041 & 0.987 & 1.236 & 3.326 & 2.28 & 0.161 & 5.457 & 1.388 \\
& NRMSE & 0.203 & 0.993 & 1.112 & 1.823 & 1.51 & 0.401 & 2.336 & 1.178 \\
& NMAE & 0.282 & 1.132 & 1.371 & 2.133 & 1.80 & 0.518 & 2.660 & 1.356 \\
Skywalking & NMSE & 0.235 & 0.264 & 0.886 & 0.202 & 0.21 & 0.143 & 0.353 & 0.875 \\
& NRMSE & 0.484 & 0.514 & 0.941 & 0.449 & 0.46 & 0.379 & 0.594 & 0.935 \\
& NMAE & 0.471 & 0.671 & 1.059 & 0.482 & 0.49 & 0.448 & 0.636 & 1.042 \\
Tensorflow & NMSE & 0.024 & 0.031 & 0.051 & 0.182 & 0.04 & 0.049 & 0.094 & 0.262 \\
& NRMSE & 0.157 & 0.176 & 0.225 & 0.426 & 0.22 & 0.223 & 0.307 & 0.512 \\
& NMAE & 0.210 & 0.250 & 0.253 & 0.429 & 0.23 & 0.250 & 0.318 & 0.526 \\
Tidb & NMSE & 0.043 & 0.124 & 6.462 & 0.23 & 0.18 & 0.546 & 0.272 & 2.858 \\
& NRMSE & 0.208 & 0.352 & 2.542 & 0.479 & 0.43 & 0.739 & 0.521 & 1.691 \\
& NMAE & 0.227 & 0.435 & 2.261 & 0.449 & 0.37 & 0.640 & 0.586 & 1.682 \\
Vscode & NMSE & 0.234 & 0.326 & 0.482 & 1.528 & 0.61 & 1.753 & 2.709 & 0.516 \\
& NRMSE & 0.484 & 0.571 & 0.694 & 1.236 & 0.78 & 1.324 & 1.646 & 0.718 \\
& NMAE & 0.435 & 0.535 & 0.559 & 1.259 & 0.75 & 1.207 & 1.516 & 0.677 \\
\bottomrule
\end{tabular}
\end{table*}

\subsubsection{Dataset}
To validate the universality and predictive accuracy of the benchmark model, this task harvested behavioral datasets from 10 highly active open-source projects throughout 2020 from OpenLeaderboard \footnote{url{https://github.com/X-lab2017/open-leaderboard}}. This list includes PyTorch, SkyWalking, TensorFlow, TiDB, VSCode, Flutter, Kibana, Kubernetes, Nixpkgs, and Rust. Some projects in this dataset contain missing behavioral data, providing a comprehensive test scenario to evaluate the benchmark model's effectiveness\cite{chen2023temporal}.

\subsubsection{Evaluation Metrics}
To assess the various models' predictive performance concerning missing and future values, NMAE, NRMSE, and NMSEA were chosen as the evaluation criteria. In practical applications, NMAE and NRMSE mainly focus on the discrepancy between predicted and actual values, while NMSEA emphasizes their relative differences. These three metrics, taken together, can comprehensively evaluate a model's predictive performance.

\subsubsection{Benchmark Result}
The OSS behavioral data with missing values from the dataset was used as the test set. Then, TRMF, Regularized MF (RMF), MF, Non-negative MF (NMF), Probabilistic MF (PMF), Basic-SVD, and BSMF were chosen as comparison algorithms. Initially, the number of iterations for the seven algorithms was set at a fixed value of 1000 (adjustable). Then, the predictive error for all normal values, excluding the missing ones, was calculated. Eventually, each algorithm's NMSE and NMAE values across the five datasets were derived.

From Table 2, it can be inferred that TAMF offers the highest completion accuracy among all the methods, confirming its effectiveness in imputing missing values. Compared to various MFs and TRMF, TAMF boasts superior completion accuracy. This result indicates that the decomposition scheme employed in the TAMF model offers significant advantages in improving the accuracy of missing value completion. TRMF comes closest to TAMF in terms of completion performance, but its overall capability is still inferior to TAMF. Hence, precise trend and cyclical feature mining of behavioral data can enhance the accuracy of missing value predictions.

\subsection{OSS Bot Identification and Classification}

\subsubsection{Dataset}

For this task, we selected the most active repositories from the GHTorrent dataset between March 2021 and March 2022 to ensure generalizability and accuracy. Accounts with over 100 log entries were identified as the "Active Account" dataset. A subset of accounts were randomly chosen from the global accounts to form the "Random Account" dataset. To expand the dataset and ensure the credibility of comparative experiments with the BIMAN and BoDeGha algorithms, we processed the data from BIMAN and BoDegHa to obtain their account's GitHub IDs. We selected active accounts from the past year (with more than 10 log entries). The data from "Active Account", "Random Account", "BIMAN Account", and "BoDeGha Account" were then merged into one dataset called the "Mixed Account". Subsequently, the "Mixed Account" data were cleaned, selecting 17 relevant features to ensure data comprehensiveness, forming the OSS bot classification dataset\cite{bi2023bothawk}.

\subsubsection{Evaluation Metrics}

To assess the performance of the bot identification model, this section employs a range of standard machine learning evaluation metrics, including Accuracy, Precision, Recall, F1-score, and AUC (Area Under the ROC Curve). Accuracy, precision, and recall mainly evaluate the model's classification capability, i.e., the model's ability to identify positive and negative classes correctly. The F1-score is a comprehensive metric of precision and recall, balancing their weights to a certain extent. The AUC reflects the model's performance when faced with different classification thresholds.

\subsubsection{Benchmark Result}

Various machine learning models were employed for the OSS bot identification task, including Logistic Regression, Decision Tree Classifier, Support Vector Machine (SVM), Gaussian Naive Bayes, K-Nearest Neighbors, Random Forest Classifier, as well as specifically designed models for the OSS bot identification task like BotHunter, BoDeGha, and BotHawk.

From Table 3, it can be observed that the BotHawk model outperforms other exemplary models, displaying the best performance across five metrics: Accuracy (0.8799), Precision (0.8930), Recall (0.8714), F1-score (0.8821), and AUC (0.9472).

\begin{table*}
\centering
\captionsetup{justification=centering}
\caption{Classifying results of nine algorithms for OSS bots}\label{tbl3}
\begin{tabular}{@{} llllll@{} }
\toprule
Model & Accuracy & Precision & Recall & F1-score & AUC \\
\midrule
LogisticRegression & 0.9234 & 0.4427 & 0.5376 & 0.4856 & 0.5760 \\
DecisionTree & 0.7995 & 0.2188 & 0.7707 & 0.3408 & 0.5024 \\
SVM & 0.8936 & 0.3495 & 0.6767 & 0.4609 & 0.5414 \\
GaussianNB & 0.9548 & 0.7514 & 0.4887 & 0.5923 & 0.5319 \\
KNeighborsClassifier & 0.8309 & 0.2472 & 0.7406 & 0.3706 & 0.5119 \\
RandomForest & 0.8817 & 0.3441 & 0.8383 & 0.4880 & 0.6486 \\
BotHunter & 0.8649 & 0.2200 & 0.7528 & 0.3405 & 0.5512 \\
BoDeGHa & 0.8286 & 0.2354 & 0.7910 & 0.3628 & 0.5049 \\
BotHawk & 0.8799 & 0.8930 & 0.8715 & 0.8821 & 0.9472 \\
\bottomrule
\end{tabular}
\end{table*}

\subsection{Project Recommendation Based on Link Prediction}
\subsubsection{Dataset}

This section introduces a benchmark dataset for evaluating open-source project recommendations, focusing on project activity and collaboration metrics from 2020. Post-processing yielded three weighted datasets: \textit{Repo\_topic}, \textit{Repo\_relation}, and \textit{Repo\_relation\_topic}. \textit{Repo\_topic} relates projects with shared tags, \textit{Repo\_relation} reflects developer contributions across projects, and \textit{Repo\_relation\_topic} combines features from both\cite{wang2022}.

\subsubsection{Evaluation Metrics}
The primary metrics for assessing link prediction algorithms are the AUC, which measures prediction accuracy, and the execution time, gauging computational efficiency.

\subsubsection{Benchmark Result}

Four network embedding-based algorithms - NodezVee, Attri2vec, GraphSAGE, and GCN - were compared against algorithms based on node local information like RA, IRA, WRA, and WICRA. Results, shown in Figure 5, indicate that local information-based algorithms generally outperform embedding based ones in terms of AUC, with RA standing out for its efficiency and accuracy. As network scale increases, the execution time for algorithms also rises proportionally.

\begin{table*}
    \centering
    \captionsetup{justification=centering}
    \caption{Results of Eight Weighted \textit{Link} Prediction Algorithms}\label{tab:results}
    \begin{tabular}{@{} lllllll@{} }
        \toprule
        Model & \begin{tabular}[c]{@{}c@{}}Repo topic \\ AUC\end{tabular} & \begin{tabular}[c]{@{}c@{}}Repo topic \\ Time(s)\end{tabular} & \begin{tabular}[c]{@{}c@{}}Repo relation \\ AUC\end{tabular} & \begin{tabular}[c]{@{}c@{}}Repo relation \\ Time(s)\end{tabular} & \begin{tabular}[c]{@{}c@{}} Repo relation \\ topic AUC \end{tabular} & \begin{tabular}[c]{@{}c@{}}  Repo relation \\ topic Time(s) \end{tabular} \\
        \midrule
        Node2Vec & 0.9475 & 1033.8656 & 0.9220 & 2373.1777 & 0.9239 & 2193.7499 \\
        Attri2Vec & 0.8513 & 403.2008 & 0.8806 & 497.9778 & 0.8806 & 547.8489 \\
        GraphSAGE & 0.9514 & 3614.3415 & 0.8986 & 4779.3441 & 0.8956 & 4945.2739 \\
        GCN & 0.9449 & 4309.6158 & 0.9064 & 11396.1526 & 0.9006 & 11460.5472 \\
        RA & 0.9593 & 119.3958 & 0.9676 & 394.6628 & 0.9692 & 392.3663 \\
        IRA & 0.9618 & 138.4095 & 0.9715 & 1240.5222 & 0.9718 & 1236.6029 \\
        WRA & 0.9609 & 126.3657 & 0.9727 & 1855.0142 & 0.9731 & 1755.1089 \\
        WICRA & 0.9676 & 161.4153 & 0.9759 & 2561.7305 & 0.9755 & 2626.3151 \\
        \bottomrule
    \end{tabular}
\end{table*}

\section{Benchmark Categories: Indices and Benchmarks and their Applications}
\subsection{Indices Category}
\subsubsection{Activity Index}

Determining whether a project remains active over the long term is crucial for developers when making technology choices, selecting components, and deciding whether to participate in a project as a contributor. OpenPerf has adopted OpenActivity, which calculates project activity by aggregating and weighting developer collaboration data\cite{zhang2020companies}. OpenActivity serves as a standardized metric to quantify a project's activity.

In this section, a statistical analysis and ranking were conducted on globally active projects on GitHub as of June 2023, using the number of participants and log increments as indicators of project activity. Here, the number of participants refers to the count of developers contributing to the project in a given month, while the log increment represents the total volume of logs for that project during the same month.

Table 5 suggests that, although \textit{pytorch/pytorch} has the highest log increment, its OpenActivity ranks only sixth. Meanwhile, \textit{NixOS/nixpkgs} witnessed fewer developers participating in that month but ranked first in OpenActivity. The overall ranking results show significant differences across different metrics. This discrepancy arises because OpenActivity, distinct from the other two metrics, is derived from the weighted sum of developer collaboration data. OpenPerf has provided statistical results for the top 10 projects on GitHub based on OpenActivity in June 2023. This offers a reference for other developers when proposing new activity index metrics, allowing them to compare and validate the rationality of their indices against these benchmark results.

\begin{table*}
    \centering
    \captionsetup{justification=centering}
    \caption{OpenActivity Ranking Comparison Results}\label{tab:openactivity}
    \begin{tabular}{@{} llll@{} }
        \toprule
        Repository & Participants & Log Increment & OpenActivity \\
        \midrule
        NixOS/nixpkgs & 1908 & 33956 & 5163.91 \\
        home-assistant/core & 3384 & 18402 & 4380.23 \\
        microsoft/vscode & 3557 & 16907 & 3643.1 \\
        flutter/flutter & 2649 & 15300 & 2938.33 \\
        MicrosoftDocs/azure-docs & 1774 & 9944 & 2884.33 \\
        pytorch/pytorch & 2102 & 52636 & 2839.17 \\
        odoo/odoo & 1285 & 34123 & 2470.15 \\
        dotnet/runtime & 862 & 16641 & 2298.13 \\
        godotengine/godot & 2247 & 11204 & 2114.51 \\
        microsoft/winget-pkgs & 539 & 26600 & 1703.35 \\
        \bottomrule
    \end{tabular}
\end{table*}

\subsection{Influence Index}
The influence of an open-source project depends on various factors. Evaluating the influence of open-source projects not only aids developers and organizations in determining which projects are worth their efforts but also offers targeted suggestions for optimizing and improving these projects. OpenPerf employs a weighted PageRank algorithm, termed as \textit{OpenRank}, to calculate the influence of a project, using it as a benchmark unit to quantify its influence.

This section presents statistics and rankings for globally active projects on GitHub as of June 2023. We compared classic degree centrality and PageRank algorithms with \textit{OpenRank}. Table 6 lists the comparison results of the top 10 projects ranked by \textit{OpenRank}.

From Table 6, the project \textit{MicrosoftDocs/azure-docs} has the highest values for degree centrality and PageRank. However, its \textit{OpenRank} is relatively lower than other projects. The project \textit{home-assistant/core} ranks first in \textit{OpenRank}. As \textit{OpenRank} calculates the centrality of a project using a weighted PageRank algorithm, its computed values tend to be higher than other metrics. This algorithm considers the impact of different collaborative behaviors on a project, hence the ranking outcome differs from other metrics. OpenPerf provides statistical results for the top 10 projects ranked by \textit{OpenRank} on GitHub for June 2023, facilitating other developers in comparing with the current three influence metrics when proposing new influence index metrics for projects.

\begin{table*}
    \centering
    \captionsetup{justification=centering}
    \caption{OpenActivity Ranking Comparison Results}\label{tab:OpenActivity}
    \begin{tabular}{@{} llll@{} }
        \toprule
        Repository & Degree Centrality & PageRank & OpenRank \\
        \midrule
        home-assistant/core & 0.015660 & 0.0035 & 2393.86 \\
        NixOS/nixpkgs & 0.008743 & 0.0008 & 2207.5 \\
        microsoft/vscode & 0.015247 & 0.003 & 1960.39 \\
        flutter/flutter & 0.012138 & 0.002 & 1460.34 \\
        pytorch/pytorch & 0.009624 & 0.0012 & 1421.18 \\
        MicrosoftDocs/azure-docs & 0.239616 & 0.08 & 1216.01 \\
        dotnet/runtime & 0.004141 & 0.0006 & 1181.12 \\
        microsoft/winget-pkgs & 0.061954 & 0.0075 & 1106.3 \\
        godotengine/godot & 0.203330 & 0.045 & 1105.51 \\
        odoo/odoo & 0.175534 & 0.043 & 907.97 \\
        \bottomrule
    \end{tabular}
\end{table*}

\begin{figure}
	\centering
		\includegraphics[scale=.15]{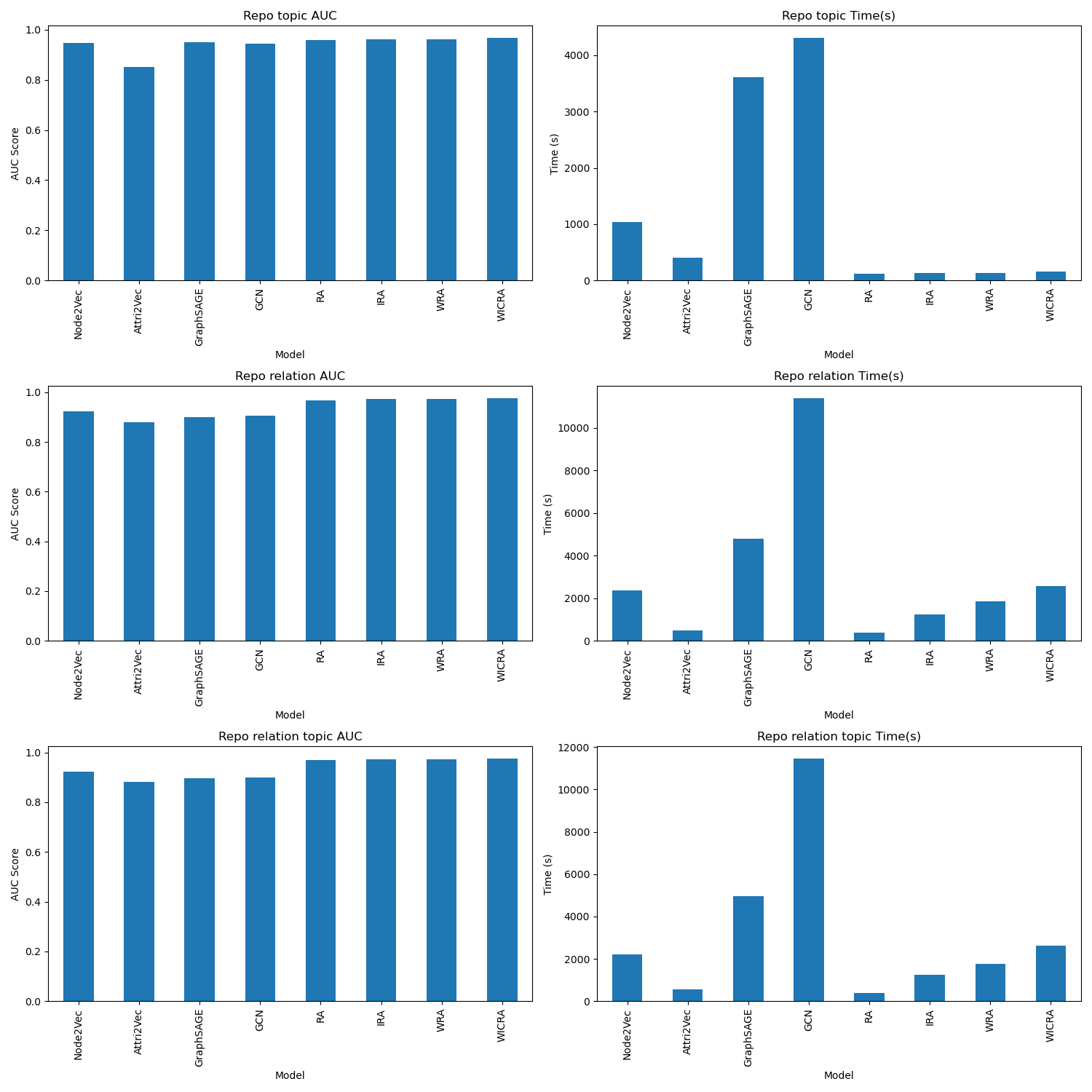}
	\caption{Results of Eight Weighted Link Prediction Algorithms}
	\label{FIG:results_of_eight_weighted}
\end{figure}

\section{Benchmark Class: OpenLeaderboard Ranking}

\textit{OpenLeaderboard} \footnote{\url{https://open-leaderboard.x-lab.info/}} is an open-source tool launched by the School of Data Science and Engineering at East China Normal University, which was officially released at the North America Open-source Summit in June 2022. This project aims to provide insights into the world of open-ource, enabling users to access a variety of information about open-source projects effortlessly. Users can view the rankings and popularity of projects and gain insight into the contributions and standing of related enterprises in the open-source domain, as depicted in Figure~\ref{FIG:results_of_eight_weighted}.

This leaderboard utilizes influence and activity index benchmarks from \textit{OpenPerf}. These index benchmarks are further defined and yield final data through tasks like "Open-source Project Influence Ranking". Additionally, top-ranked entities (e.g., Top 100) are considered as benchmark sets.

At present, \textit{OpenLeaderboard} has become a barometer in the open-source domain of the software industry. It has been well-applied in various areas, including open-source component selection, global open-source situation analysis, and observation of enterprises' interactions with the open-source community.

\begin{figure}
	\centering
		\includegraphics[scale=.15]{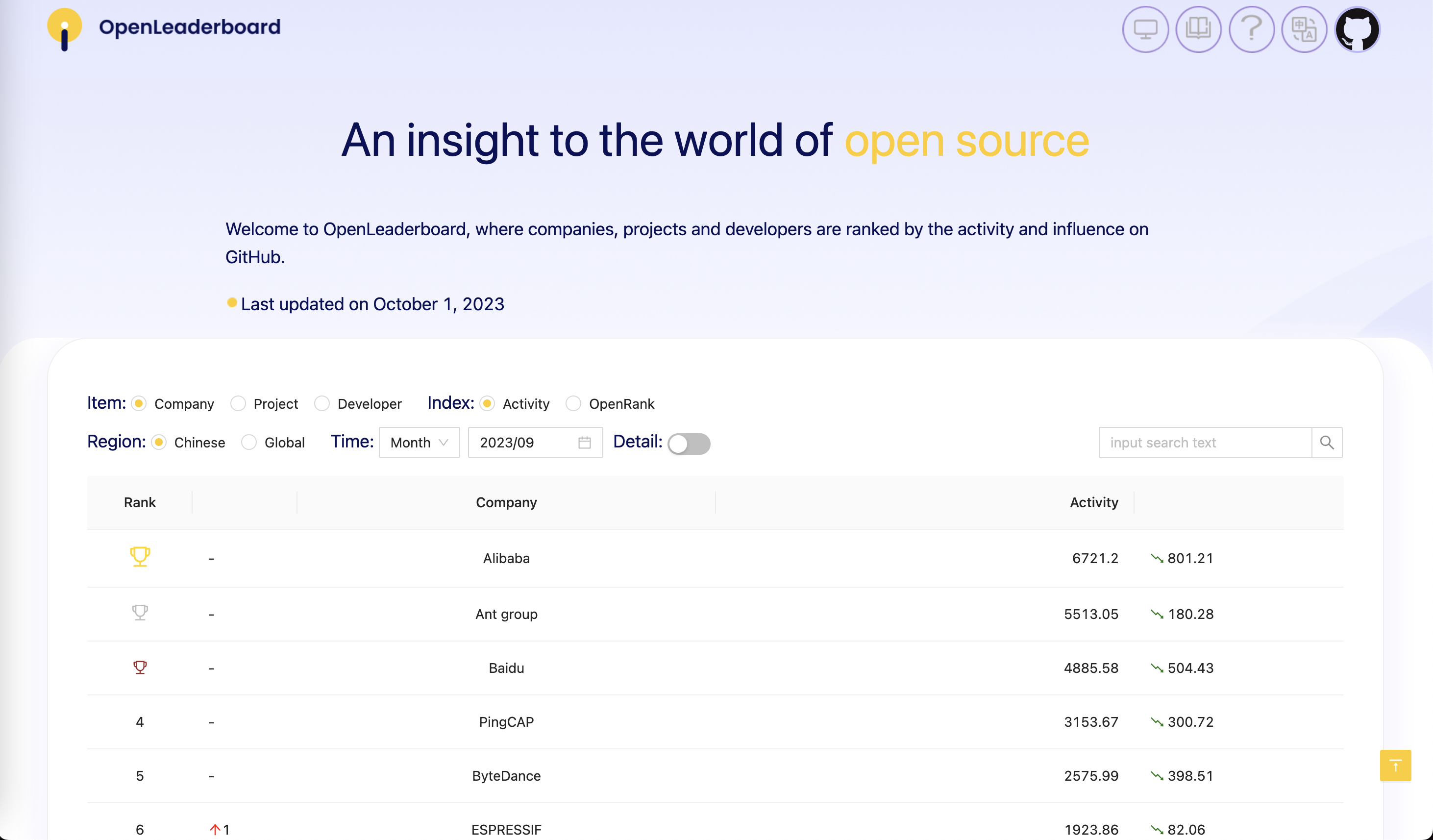}
	\caption{OpenLeaderBoard Project Showcase}
	\label{FIG:openleaderboard_project_showcase}
\end{figure}

\subsection{Industry Application 1: Ant Group OSPO Open-source Governance Dashboard}

Corporate open-source governance dashboard to the open-source software utilized by the enterprise and collaborations inside and outside the firm. This includes aspects like software selection and usage, management and maintenance, collaborations with external businesses and communities, and decisions regarding initiating one's own open-source projects for deriving benefits. It also encompasses adopting open-source community collaboration models to enhance internal efficiency and quality. Domestic and international IT companies' Open-source Program Offices (OSPO) are increasingly adopting best practices that combine corporate open-source governance processes with data visualization dashboards. \textit{OpenPerf} is progressively becoming the tool of choice for many enterprises.

\begin{figure}
	\centering
		\includegraphics[scale=.20]{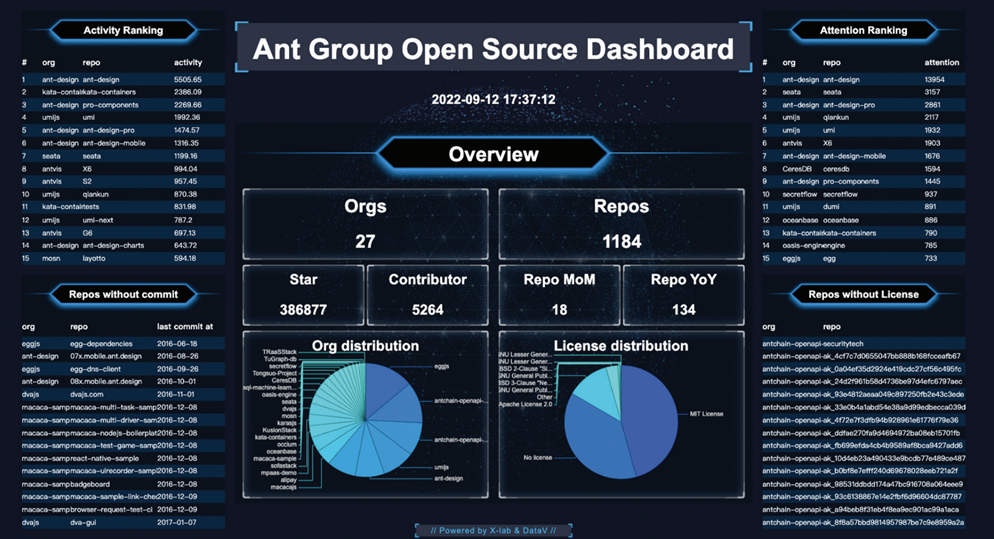}
	\caption{Ant Group Open-source Dashboard}
	\label{FIG:ant_group_dashboard}
\end{figure}

Ant Group's OSPO Open-source Governance Dashboard\cite{xia2023lessons} employs influence, activity, and risk index benchmarks from \textit{OpenPerf}. These benchmarks are further defined and yield final data through tasks such as "Open-source Software Supply Chain Risk Prediction" and "Anomaly Detection in Open-source Communities". This benchmarking service holistically analyzes the development status of multiple open-source projects within the group. It effectively quantifies the activity and influence of different projects, offering insights for their sustainable and healthy growth.

\subsection{Industry Application 2: Alibaba Open-source Developer Contribution Incentive Ranking}

The objective of assessing individual developer influence is to calculate the impact of all developers in the global open-source ecosystem based on their collaborative efforts across diverse projects. Precisely evaluating individual contributions and influence aligns developers' efforts with rewards, promoting sustained open-source collaboration and the scalable growth of open-source communities. Numerous enterprises both domestically and internationally utilize the developer influence benchmark proposed in \textit{OpenPerf} to evaluate the influence of developers in corporate open-source projects. This further shapes the incentives for developers in corporate open-source projects, encouraging more newcomers to participate in open-source project development, as depicted in Figure 8.

\begin{figure}
	\centering
		\includegraphics[scale=.20]{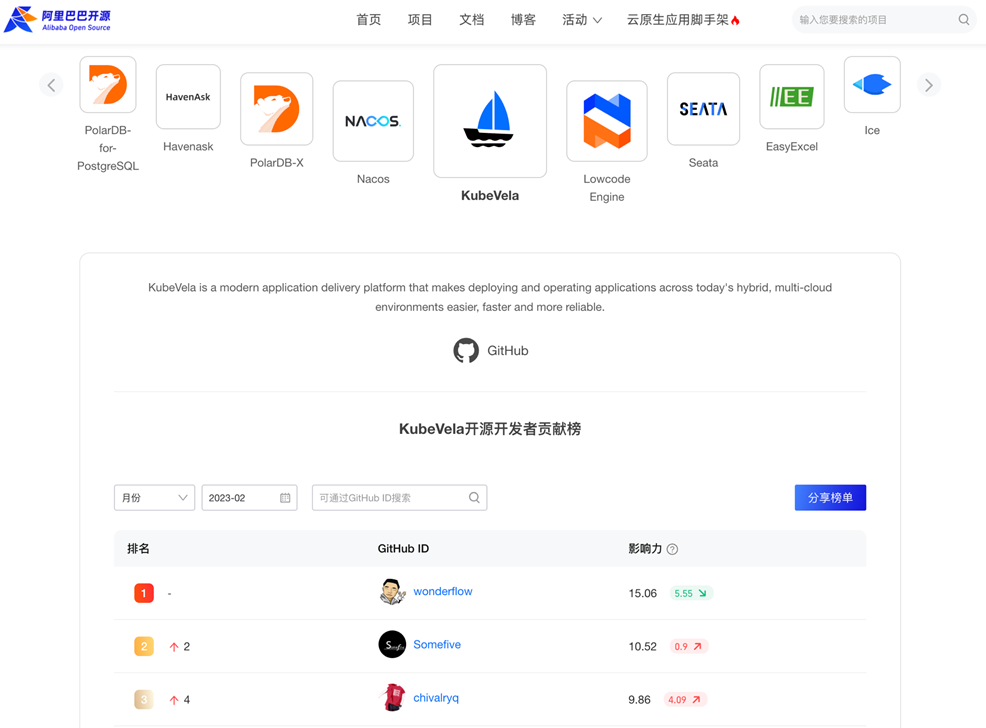}
	\caption{Alibaba Open Source Contribution Leaderboard}
	\label{Contribution-Leaderboard}
\end{figure}

Using Alibaba's Open-source Developer Contribution Leaderboard\footnote{\url{https://opensource.alibaba.com/contribution_leaderboard/details?projectValue=sealer&timeType=month&time=1685548800000}} as an example, the leaderboard employs influence and activity indices from \textit{OpenPerf}. These indices are defined through "Open-source Project Influence Ranking" tasks to produce the final data. This leaderboard captures developer influence rankings for various projects. Alibaba grants rewards to developers based on these influence scores. Consequently, a significant rise is observed in the number of developers willing to communicate on the GitHub platform, with marked improvements in issues, PRs, likes, and especially in issue comment counts. This highlights that a rational evaluation of open-source developer influence can incentivize and further promote the healthy development of the open-source ecosystem.

\subsection{Industry Application 3: Procedural Evaluation in Open-source Software Courses}

Sustainable development of the open-source ecosystem requires support from open-source talents; education in this field is pivotal. Numerous universities domestically and abroad have introduced open-source-related courses. East China Normal University, starting from 2019, has gradually offered open-source courses aimed at graduate students, computer science undergraduates, and general university-wide courses. The course evaluation process integrates an open-source approach.

\begin{figure}
	\centering
		\includegraphics[scale=.20]{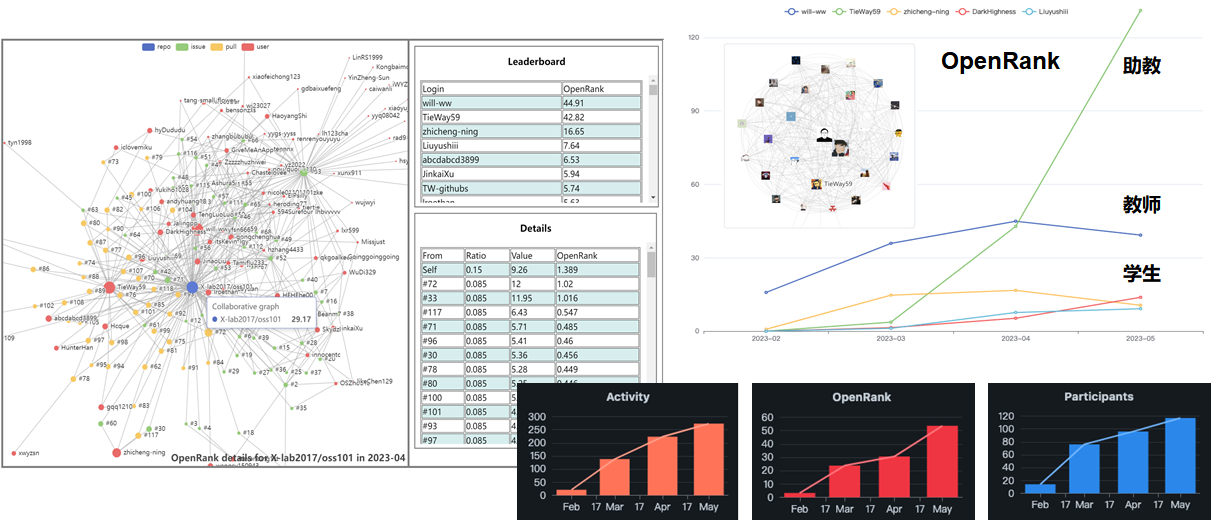}
	\caption{Analysis of Student Influence in Open-source General Education Courses}
	\label{FIG:student_influence}
\end{figure}

\textit{Open-source Software General Knowledge}\footnote{\url{https://github.com/X-lab2017/oss101/}} is a general course intended for all university students. The learning process, assignments, and hands-on training are all conducted within code repositories. The course team employs activity and influence benchmarks from \textit{OpenPerf} for procedural evaluation of students, yielding favorable outcomes. This not only spurs students to actively engage in repository-based open-source activities and drives open-source project development but also enables students to grasp the essence of open-source collaboration deeply, laying a foundation for nurturing exceptional open-source talents.

\section{Conclusion}
In recent years, with the development of open-source software, related research has gradually emerged as an academic focus. Against this backdrop, this paper delves into the core scientific and engineering principles of benchmark testing, establishing a benchmark—\textit{OpenPerf}, aimed at fostering the sustainable growth of the open-source ecosystem. This benchmark defines 9 data science task benchmarks in the open-source realm and further implements 3 task benchmarks, 2 index benchmarks, and 1 standard benchmark. It strives to provide researchers with a robust and scientific tool and platform, ultimately promoting open-source research and practice integration and advancement.

\textit{OpenPerf} emphasizes the standardization and regularization of data science in the open-source. It is anticipated to evolve, offering more abundant and precise benchmarking data and results tailored to various sub-domains in open-source research, such as project health, maturity, and community participation.

\textit{OpenPerf} successfully bridges academia and practical application, inspiring numerous high-quality academic papers and providing continuous services to the academic community, industry, foundations, and other organizations. Notably, the China Electronics Technology Standardization Institute adopted two of our benchmark indices, serving as essential reference standards for evaluating open-source community governance.

\bibliographystyle{cas-model2-names}

\bibliography{cas-dc-template}





\end{document}